\documentclass[aps,reprint,superscriptaddress]{revtex4-1}

\usepackage{amssymb}
\usepackage{amsmath}
\usepackage{amscd}
\usepackage{latexsym}
\usepackage{graphicx}

\usepackage{enumitem}

\newcommand{\be}{\begin{equation}}
\newcommand{\ee}{\end{equation}}

\newcommand{\dlt}{\delta}

\newcommand{\br}{{\bf r}}
\newcommand{\bd}{{\bf d}}
\newcommand{\bfe}{{\bf e}}

\newcommand{\bP}{{\bf P}}

\newcommand{\bE}{{\bf E}}

\newcommand{\Om}{\Omega}

\newcommand{\ra}{\rightarrow}

\newcommand{\dgr}{\dagger}

\begin{document}

\title{On ultrafast polarization switching in ferroelectrics}

\author{V.I. Yukalov}

\affiliation{Bogolubov Laboratory of Theoretical Physics,
Joint Institute for Nuclear Research, Dubna 141980, Russia}

\affiliation{
Instituto de Fisica de S\~{a}o Carlos, Universidade de S\~{a}o Paulo,
CP 369, S\~{a}o Carlos 13560-970, S\~{a}o Paulo, Brazil}

\author{E.P. Yukalova}

\affiliation{Laboratory of Information Technologies,
Joint Institute for Nuclear Research, Dubna 141980, Russia}

\begin{abstract}   

Recently, a method of ultrafast polarization switching in ferroelectrics has been 
suggested. The basic idea of the method is to employ the effect of self-acceleration 
of polarization dynamics due to a resonator feedback field. This is the idea of principle 
whose efficiency is demonstrated by an Ising-type model in a transverse field. Of course, 
the practical realization of the method requires the choice of appropriate materials, 
which is a separate problem. For example, the standard order-disorder ferroelectrics with 
spatially symmetric double-well potentials cannot be used for this purpose, since, 
because of the symmetry, they lack the transverse polarization. However, ferroelectrics
with asymmetric potentials, possessing this polarization, can be used. Moreover, if the 
potential asymmetry, hence the transverse polarization, could be regulated, e.g., by shear 
stress or shear strain, this could provide a tool for governing the process of polarization 
switching. 
  
\end{abstract}

\maketitle

The main goal of paper \cite{Yukalov_1} has been to attract attention to the possibility 
of accelerating the polarization switching in ferroelectrics by using the self-acceleration 
effect caused by the action of a resonator-cavity feedback field. This is the idea of 
principle that, to our knowledge, has not been considered for ferroelectrics before. 
It goes without saying that the method is not necessarily applicable to any particular 
material. Thus in Ref. \cite{Abalmasov_2}, it is mentioned that the standard order-disorder 
ferroelectrics with spatially symmetric double wells cannot be used for this purpose. 
Below we explain that the model considered in Ref. \cite{Yukalov_1} assumes the general 
case of asymmetric potentials for which the method is applicable. We also emphasize 
that, since there are various types of ferroelectric systems 
\cite{Blinc_3,Lines_4,Strukov_5,Whatmore_6}, there can exist other materials for which 
the suggested idea could work.  

Of course, to illustrate the idea, one has to consider some model. We considered an 
Ising-type model in a transverse field. This kind of models, employing the spin 
representation, is widely used for order-disorder ferroelectrics 
\cite{Blinc_3,Lines_4,Strukov_5,Whatmore_6} and relaxor ferroelectrics \cite{Hong_7}.

In order that the suggested method could be realized, the existence of two spin
components of polarization, longitudinal and transverse, is required. If one keeps 
in mind an order-disorder ferroelectric with lattice-site double wells that are 
ideally symmetric with respect to spatial inversion (especially with respect to the
inversion $x \ra -x$), then there is only a longitudinal component, and the sample 
polarization is expressed through the $z$-component of the spin operator. But in the 
general case of an asymmetric potential, the sample polarization contains a term with 
the $x$-component of spin. The asymmetry can be induced by shear stress or strain, or 
by incorporating into the sample admixtures or vacancies. Thus, the inclusion of 
vacancies in order-disorder ferroelectrics is attributed to the breaking of spatial 
inversion symmetry along different directions \cite{Yoo_8}. The symmetry in order-disorder 
ferroelectics can be distorted by the action of a transverse electric field \cite{Fugiel_9}. 
 
In the paper \cite{Yukalov_1}, the general case of an asymmetric site potential 
is treated, when the expression for polarization possesses both spin components, 
longitudinal and transverse. In order to be explicit, let us briefly show how these 
components arise.

The most general approach requires to start from the microscopic Hamiltonian in the 
second-quantization representation  
$$
\hat H = \int \psi^\dgr(\br) H_1(\br) \psi(\br) \; d\br \; - \; 
\hat \bP \cdot \bE_{tot} \; +
$$
\be
\label{1}
+ \; 
\frac{1}{2} \int \psi^\dgr(\br) \psi^\dgr(\br') 
\Phi(\br-\br') \psi(\br')\psi(\br) \; d\br d\br' \; .
\ee
Here $\psi$ are field operators, the site Hamiltonian is
\be
\label{2}
H_1(\br) = -\; \frac{\nabla^2}{2m} + U(\br) \; ,
\ee
the potential $U({\bf r})$, generally, is not symmetric with respect to the spatial 
inversion ${\bf r} \ra - {\bf r}$, $\Phi$ is the interaction potential, $E_{tot}$ 
is an external electric field acting on the polarization operator
\be
\label{3}
\hat \bP = \int \psi^\dgr(\br) \bP(\br) \psi(\br) \; d\br \;   ,
\ee
where the local polarization $\bP(\br)$ is caused by the charge distribution 
satisfying the condition of the sample neutrality. 

Considering an insulating sample, where particles are localized in the vicinity of 
the lattice sites, the field operators can be expanded over localized orbitals, for 
example over the well-localized Wannier functions \cite{Marzari_10},
\be
\label{4}
 \psi(\br) = \sum_{nj} c_{nj} w_n(\br-\br_j) \;  .
\ee
For the unity filling factor, the no-double-occupancy condition is valid:
$\sum_n c_{nj}^\dgr c_{nj} = 1$ and $c_{mj} c_{nj} = 0$. For an insulating 
lattice, the no-hopping condition is satisfied: $c_{mi}^\dgr c_{nj} = 
\dlt_{ij} c_{mj}^\dgr c_{nj}$.

One assumes that the most populated are the lowest two energy levels, while the 
other levels can be neglected. This allows one to introduce the spin representation
$$
 c_{1j}^\dgr c_{1j} = \frac{1}{2} \; + \; S_j^x \; , \qquad 
c_{2j}^\dgr c_{2j} = \frac{1}{2} \; - \; S_j^x \; ,
$$
\be
\label{7}
c_{1j}^\dgr c_{2j} =  S_j^z - i S_j^y \; , \qquad  
 c_{2j}^\dgr c_{1j} =  S_j^z + i S_j^y \; .
\ee
Thus, omitting nonoperator terms, we come to the Hamiltonian
$$
 \hat H = - \sum_j \left( \Om_j S_j^x - H_j S_j^z\right) \; - \;
\sum_j \hat \bP_j \cdot \bE_{tot} \; +
$$
\be
\label{8}
 + \; 
\frac{1}{2} \sum_{i\neq j} \left( B_{ij}S_i^x S_j^x - J_{ij} S_i^z S_j^z \right) \; ,
\ee
in which $\Om_j \equiv H_{jj}^{22} - H_{jj}^{11} + \frac{1}{2} \sum_i C_{ij}$, 
$H_j \equiv H_{jj}^{12} + H_{jj}^{21}$, 
$B_{ij} \equiv V_{ij}^{1111} + V_{ij}^{2222} - 2 V_{ij}^{1221}$, 
$C_{ij} \equiv V_{ij}^{2222} - V_{ij}^{1111}$, and $J_{ij} \equiv - 4V_{ij}^{1122}$,
with $H_{ij}^{mn}$ being the matrix elements of Hamiltonian (\ref{2}) over the 
Wannier functions and $V_{ij}^{mnkl}$, the corresponding matrix elements of the 
interaction potential. The polarization operator takes the form
\be
\label{9}   
\hat \bP_j \equiv \bd_0 S_j^z + \bd_1 S_j^x \;   ,
\ee
where $\bd_0 \equiv \bP_{12} + \bP_{21}$, $\bd_1 \equiv \bP_{11} - \bP_{22}$, and
$$
\bP_{mn} \equiv \int w_m^*(\br-\br_j) \bP(\br) w_n(\br-\br_j) \; d\br \; .
$$
The total external electric field contains a longitudinal and a transverse components,
$\bE_{tot}= E_0 \bfe_z + E \bfe_x$. Here $E_0$ is a fixed external field and $E$ is 
the feedback field of a resonant cavity. 
  
The term $H_j$ can be included into the field $E_0$. The magnitude of the interaction
parameter $B_{ij}$ is usually much smaller than the tunneling frequency $\Omega_j$ that 
can be taken the same for all lattice sites, $\Omega_j = \Omega$. Omitting the term with 
$B_{ij}$ is not principal, since it is easy to show \cite{Yukalov_19} that its main role 
is the renormalization of the frequency $\Omega_j$. As a result, we obtain the Hamiltonian
$$
\hat H = - \Om \sum_j S_j^x \; - \; 
\frac{1}{2} \sum_{i\neq j} J_{ij} S_i^z S_j^z \; -
$$
\be
\label{11}
- \; 
\sum_j \hat \bP_j \cdot \bE_{tot} \; .
\ee

When the potential configuration at each lattice site is symmetric with respect 
to spatial inversion (especially with respect to the inversion $x \ra -x$), so that 
the density $|w_n({\bf r})|^2$ is also symmetric with respect to the spatial inversion, 
then the diagonal matrix elements ${\bf P}_{nn}$ are zero, because of which the 
polarization operator contains only the longitudinal spin component ${\bf d}_0 S_j^z$. 
However in the general case, when the potential relief is not inversion symmetric, the 
polarization operator contains both, the longitudinal as well as transverse spin 
components, as in Eq. (\ref{9}). This general case is assumed in Ref. \cite{Yukalov_1}. 

Thus, in the general situation, the polarization operator possesses two spin components, 
longitudinal and transverse. This is sufficient for realizing the effect of the 
self-acceleration of the polarization switching by the resonator feedback field, which 
is the main point of the paper \cite{Yukalov_1}. It is useful to note that, in addition 
to different ferroelectrics \cite{Blinc_3,Lines_4,Strukov_5,Whatmore_6,Hong_7} and 
multiferroics \cite{Wang_11,Xiang_12}, ferroelectric-type spin models are widely used for 
describing the systems of polar molecules, Rydberg atoms, Rydberg-dressed atoms, 
dipolar ions, vacancy centers in solids, and quantum dots 
\cite{Baranov_13,Krems_14,Dulieu_15,Baranov_16,Gadway_17,Birman_18}. These systems 
can form self-assembled lattice structures or can be loaded into external potentials 
imitating crystalline models. The characteristics of these dipolar materials can be 
varied in a very wide range. Therefore it looks feasible to find the appropriate 
material for the realization of the effect considered in the paper \cite{Yukalov_1}.
    
Moreover, the fact that the potential asymmetry leads to the appearance of the transverse
polarization component containing $S^x_j$ suggests the way of regulating the polarization 
switching. Thus in a sample with symmetric wells, where there is no transverse 
polarization, the longitudinal polarization can be frozen. By inducing the potential 
asymmetry, for instance by subjecting the material to shear stress and shear strain, which 
would induce the transverse polarization, it would be possible to trigger the switching 
process.

\end{document}